# Realizing long radio-frequency quadrupole accelerators with multiple shorter and independent cavities

Chuan Zhang [*, 1], Holger Podlech [2], Eugene Tanke [3]

[1] *GSI Helmholtz Center for Heavy Ion Research, Planckstr. 1, Darmstadt, Germany*

[2] *Institute for Applied Physics, Goethe University, Frankfurt a. M., Germany*

[3] *ESS, Lund, Sweden*

*Abstract*:

It is well known that the tuning of a long radio-frequency quadrupole accelerator is demanding. This study investigated how to realize efficient long RFQ accelerators with multiple shorter and independent cavities. From the RF point of view, the use of shorter cavities has many advantages e.g. enlarged mode separation and simplified RF tuning, but the beam matching between cavities will become an issue, especially at high current and low energy. Taking a more than 9 m long, high current machine as an example, this paper presents the design concepts and methods leading to good beam quality and high beam transmission throughout a multi-stage RFQ accelerator.

[#] e-mail: c.zhang@gsi.de

# I. INTRODUCTION

Along with the development in the last 50 years, the Radio-Frequency Quadrupole (RFQ) accelerator has become a kind of very important proton and ion accelerating structure either as stand-alone machine or as injector to a large facility. This structure is excellent in capturing, focusing and bunching a dc beam from an ion source and the low energy transport section, but is not very efficient for accelerating it to high beam energy. Typically, the output beam energy of an RFQ accelerator is lower than 3 AMeV, and its structure length is shorter than 4 m. For some projects (see Fig. 1), however, longer RFQs have been constructed or operated. Except for the ISAC RFQ [1] and the HSI ("High Current Injector" translated from German "Hochstrominjektor") RFQ [2], they are all using the 4-vane structure based on the $TE_{210}$ mode.

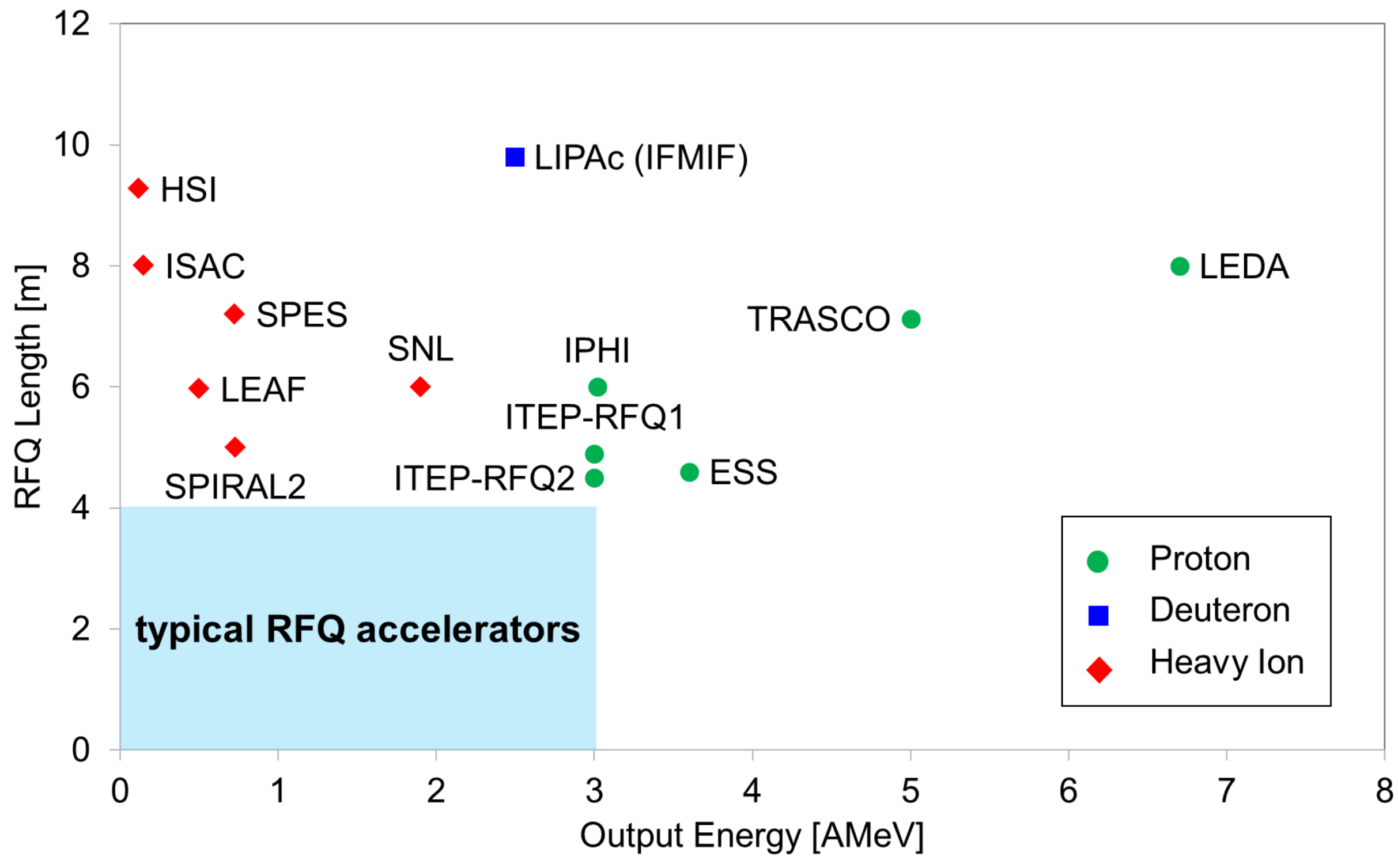

FIG. 1. Overview of some long (> 4 m) RFQ accelerators constructed or operated worldwide (note: the HSI and SPIRAL2 RFQs are also used for light ions).

Usually, two major aspects in tuning an RFQ are: 1) to keep a good frequency separation between the operating mode and the nearest unwanted mode for enhancing the longitudinal stability; 2) to stabilize the quadrupole mode from disturbing dipole components. For the $TE_{210}$ mode, the sensitivity to tuning errors can be expressed as proportional to $(L/\lambda)^2$, where $L$ is the RFQ length and $\lambda$ is the free-space wavelength [3, 4]. It can be seen that the longer the RFQ is, the more difficult the tuning is. The HSI RFQ uses the interdigital H-mode (IH) structure based on the $TE_{110}$ mode. Similar to the 4-vane structure, the magnetic field lines in the HSI RFQ are parallel to the beam axis and turning around at the two ends of the cavity.

To address the difficulties in tuning long RFQs, several solutions have been proposed and developed. One is the so-called "resonant coupling" approach which divides a long RFQ accelerator into short (e.g. 1 – 2 m) sections and couples them together by short (a few mm) gaps between the end regions [5]. The RF simulation of a 4 m long, 425 MHz RFQ shows that the frequency difference between the mode of interest and the neighboring mode can be increased from 1.9 MHz to 6.6 MHz via resonant coupling [5]. After the successful application to the 8 m long LEDA RFQ [6], this approach has been taken for many later long 4-vane RFQs e.g. TRASCO [7] and IFMIF [8]. This kind of coupled structure needs only one RF power source and one beam dynamics design, so it is still one RFQ.

Another idea is to use a so-called "Tandem-RFQ" i.e. two sequential RFQs with independent RF systems. Three examples are: 1) the constructed SNL RFQ with a total length $L_{total}$ = 6 m [9], the constructed superconducting PIAVE RFQ with $L_{total}$ = ~ 2 m [10] (it is not a long RFQ, but uses a similar idea), and the proposed normal-conducting EURISOL RFQ with $L_{total}$ = 7.8 m [11]. A challenge for using this concept is addressing the beam matching issue between the cavities, especially in the transverse phase spaces. Usually at the entrance to an RFQ accelerator, converging and similar particle distributions in both $x$ and $y$ planes are desired. But at the exit of an RFQ, typically the output beam is converging in one transverse plane and diverging in another one. The matching problem can become very critical at high beam currents. The SNL and PIAVE projects need low beam currents up to eμA, so no special care is necessary between two separated cavities. In case of EURISOL which needs to work for heavy ion applications with negligible beam currents at continuous wave but also for lighter ions with currents up to 7.5 emA at pulsed operation, a same tank with a thin RF-shielding wall has been taken for the two RFQs [11]. The 1.7 cm long gap between the two sets of electrodes allows a very quick transition.

The motivation of this study is to investigate how to realize efficient long RFQs using multiple shorter and independent cavities with not only improved field stability and easy RF tuning but also high beam transmission and good beam quality, including at high currents. To introduce the design concepts and methods for fulfilling this purpose, the 9.3 m long HSI RFQ is adopted as an example by this study.

Serving as an injector to the UNILAC, the HSI RFQ can accelerate a wide variety of particle species from protons to uranium ions in the energy range of 2.2 – 120 AkeV. Some milestones of this machine are as follows:

- In 1996: the first design of the HSI RFQ for a 16.5 emA $U^{4+}$ beam was started [12].
- In 1998: the first HSI RFQ was constructed [2] (hereafter it is called as Version-1998).
- In 1999: the Version-1998 RFQ was put into operation.
- In 2004: the electrodes were replaced with new ones (with an improved radial matching section for a bigger acceptance).
- In 2008: the second HSI RFQ was designed and produced for a 20 emA $U^{4+}$ beam [13] (hereafter it is called as Version-2008).
- In 2009: the Version-2008 RFQ was put into operation.
- From 2009 until now: the second HSI RFQ is in routine operation (in 2019, the electrodes were renewed using the same design).

TABLE I. Main design parameters of the two constructed HSI RFQs.

| | **Version-1998** | **Version-2008** |
|---|---|---|
| Frequency $f$ [MHz] | 36.136 | 36.136 |
| Input energy $W_{in}$ [AkeV] | 2.2 | 2.2 |
| Output energy $W_{out}$ [AkeV] | 120 | 120 |
| Design beam current $I$ [emA] | 16.5 | 20 |
| Inter-vane voltage $V$ [kV] | 125 | 155 |
| Mid-cell aperture $r_0$ [mm] | 5.2 – 7.8 | ~ 6.0 |
| Maximum surface E-field $E_{s,\,max}$ [MV/m] | 31.8 | 31.2 |
| Number of cells | 357 | 409 |
| Total length $L_{total}$ [m] (including the gaps between the endplates and electrodes) | 9.27 | 9.27 |
| Input transverse emittance [π mm mrad] | | |
| • Unnormalized, total emittance $\varepsilon_{in,\,unnorm.,\,total}$ | 138 | 210 |
| • Normalized, rms emittance $\varepsilon_{in,\,n.,\,rms}$ | 0.050 | 0.076 |
| Transmission $T$ [%] | 89.5 | 88.5 |

In the RF study performed for the first HSI RFQ [14], one can see that: 1) for the full-length tank, the frequency difference between the operating mode and the nearest mode is around 2 MHz; 2) for short tank segments, the higher resonant modes are by about a factor 8 above the fundamental mode.

Instead of using a non-constant mid-cell aperture $r_0$ varied from 5.2 mm to 7.8 mm like the Version-1998 RFQ (see Table I), the Version-2008 adopted an almost constant $r_0$ (~ 6.0 mm) [13] for the main RFQ, in order to keep the distributed capacitance along the accelerator nearly constant for easier tuning. This has no influence on the frequency gap for the mode separation (still ~ 2 MHz). Another remarkable change in the Version-2008 design is that the inter-vane voltage $V$ was increased from 125 kV to 155 kV [13]. On one side, it is because of a higher design beam current. On the other side, this is necessary for meeting the demand of having a constant $r_0$ for such a long machine. However, this change required more RF power (the power is proportional to $V^2$).

The design strategies for this study are as follows: 1) to increase the frequency gap for mode separation by using multiple shorter and independent cavities; 2) to apply almost constant but different $r_0$ values for different cavities, which eases tuning; 3) to adapt the inter-vane voltage $V$ individually according to the different space-charge situation in each cavity.

This study has been performed in three steps:

- The RFQ was divided into 2 pieces with drift space in between.
- The RFQ was divided into 2 pieces with a Medium Energy Beam Transport (MEBT) section in between.
- The RFQ was divided into 3 pieces with drift space in between.

The HSI RFQ covers the energy range of 2.2 – 120 AkeV. No matter a 2-piece solution or a 3-piece solution, therefore, the transition energy $W_{trans}$ between the cavities will be much lower than those of the three above-mentioned Tandem-RFQs (SNL: $W_{trans}$ = 1.22 AMeV [9]; PIAVE: $W_{trans}$ = 341.7 AkeV [10]; EURISOL: $W_{trans}$ = 260 AkeV [11]). Furthermore, the much higher design beam current, 20 emA, will make the design work more challenging, so special measures are needed.

## II. 2-RFQ SOLUTION WITH DRIFT SPACE BETWEEN CAVITIES

Firstly, a solution using two RFQ cavities with drift space in between has been investigated. Each cavity should have a structure length that is roughly half of the original RFQ length i.e. ~ 4.64 m. Because the first cavity (RFQ 1) is more important for bunching and the second one (RFQ 2) is more important for acceleration, the transition energy has been chosen as 53.6

AkeV which means that the RFQ 1 will cover an energy gain less than half of what is required for the whole HSI RFQ.

To reach good beam quality and high beam transmission, the design of the RFQ 1 is very important as well as relatively difficult, because it performs the main bunching at lower energy and sees stronger space-charge effects. Therefore, a so-called $\frac{\varepsilon_l}{\varepsilon_t}$ (ratio of the longitudinal and transverse emittances) = 1.0 design approach [15] has been adopted for the RFQ 1, as it has been successfully applied for designing a 325 MHz, 3 MeV, 70 – 100 mA proton RFQ with high beam transmission, short structure length, and high beam quality. Combined with the New Four-Section Procedure (NFSP) [16], this design approach takes advantage of the large resonance-free area provided by the $\frac{\varepsilon_l}{\varepsilon_t}$ = 1.0 condition for the beam evolution and can efficiently minimize emittance transfer and consequently reduce beam instabilities.

To achieve a good beam matching from the RFQ 1 to the RFQ 2, a 3.2 cm long transition cell with no modulation has been added in front of the exit fringe-field region. With this cell, the beam energy is unchanged, but one obtains phase-space ellipses at the RFQ output with Twiss parameter $\alpha_{\mathrm{Twiss}} \cong 0$ in both transverse directions. This will make the matching of the beam into the second cavity smoother.

Based on the above-mentioned methods, a 9.28 m long, 36.136 MHz linac with two RFQ cavities (4.64 m and 4.63 m long, respectively) and a drift space (drift length $d$ = 1 cm) in between has been designed. This solution is called as “2-RFQ with drift” in the following text. The evolution of the main design parameters along the accelerating channel is shown in Fig. 2, where $a_{\mathrm{min}}$ is the minimum electrode aperture, $m$ is the electrode modulation, $\varphi_s$ is the synchronous phase, $V$ is the inter-vane voltage, and $B$ is the transverse focusing strength, respectively. It can be seen that: 1) the $V$ values have been set differently for the two cavities;

2) in the second cavity, all parameters are held almost constant, except the synchronous phase has been changed shortly at the entrance and at the exit for better matching.

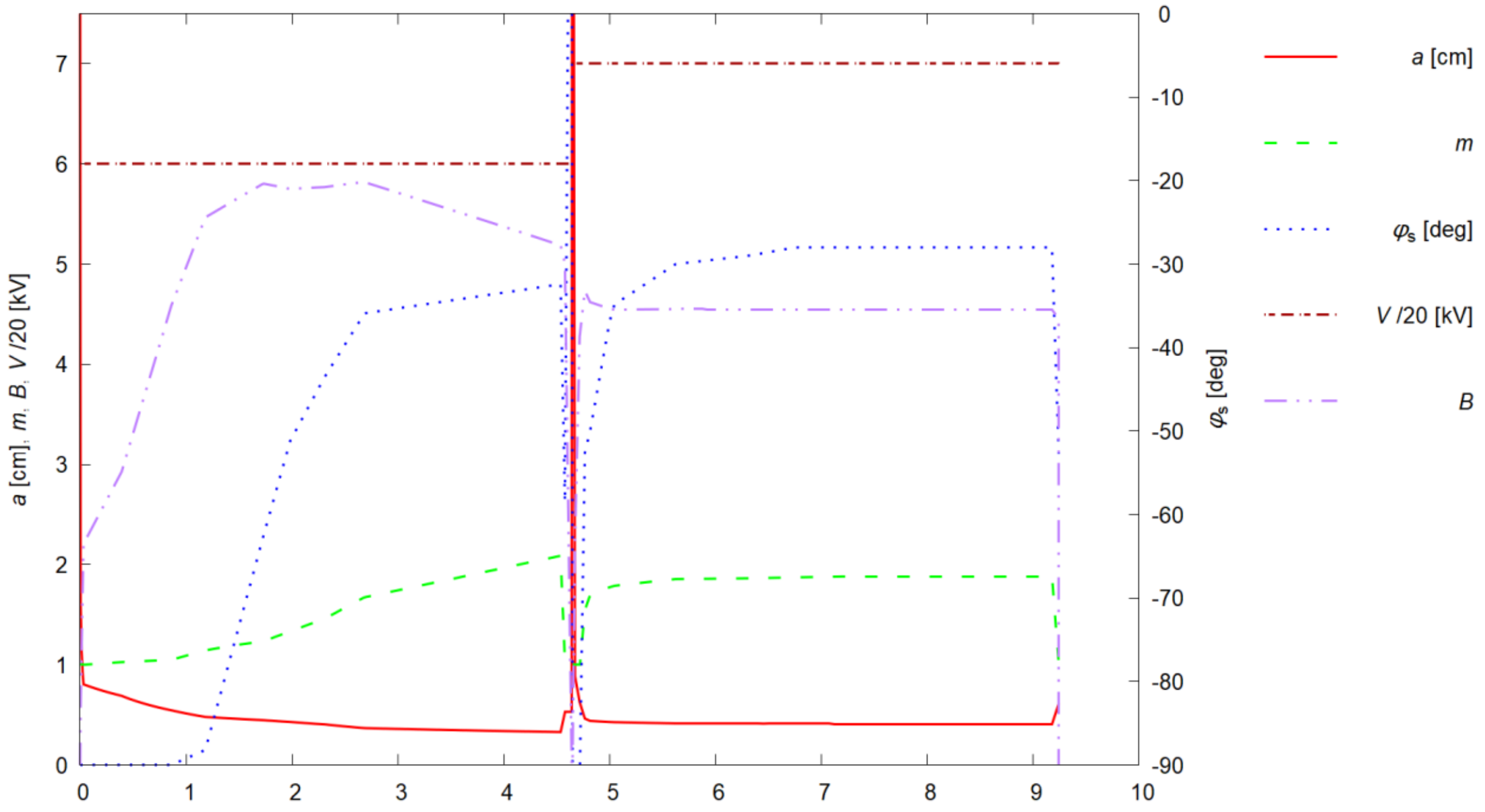

FIG. 2. Main design parameters of the "2-RFQ with drift" solution.

A comparison of the mid-cell aperture $r_0$ values between the new solution and the two existing designs is given in Fig. 3. Compared to the Version-1998 and Version-2008 RFQs, the 2-RFQ solution has a much bigger $r_0$ at the entrance, which helps increasing the transverse acceptance as well as the beam transmission. Then its $r_0$ value becomes smaller than those of the other two designs, which supports using lower $V$ and saving RF power. Last but not least, the $r_0$ value is quasi-constant in the main part of the first cavity as well as in the whole second cavity, which is favorable for an easy RF tuning.

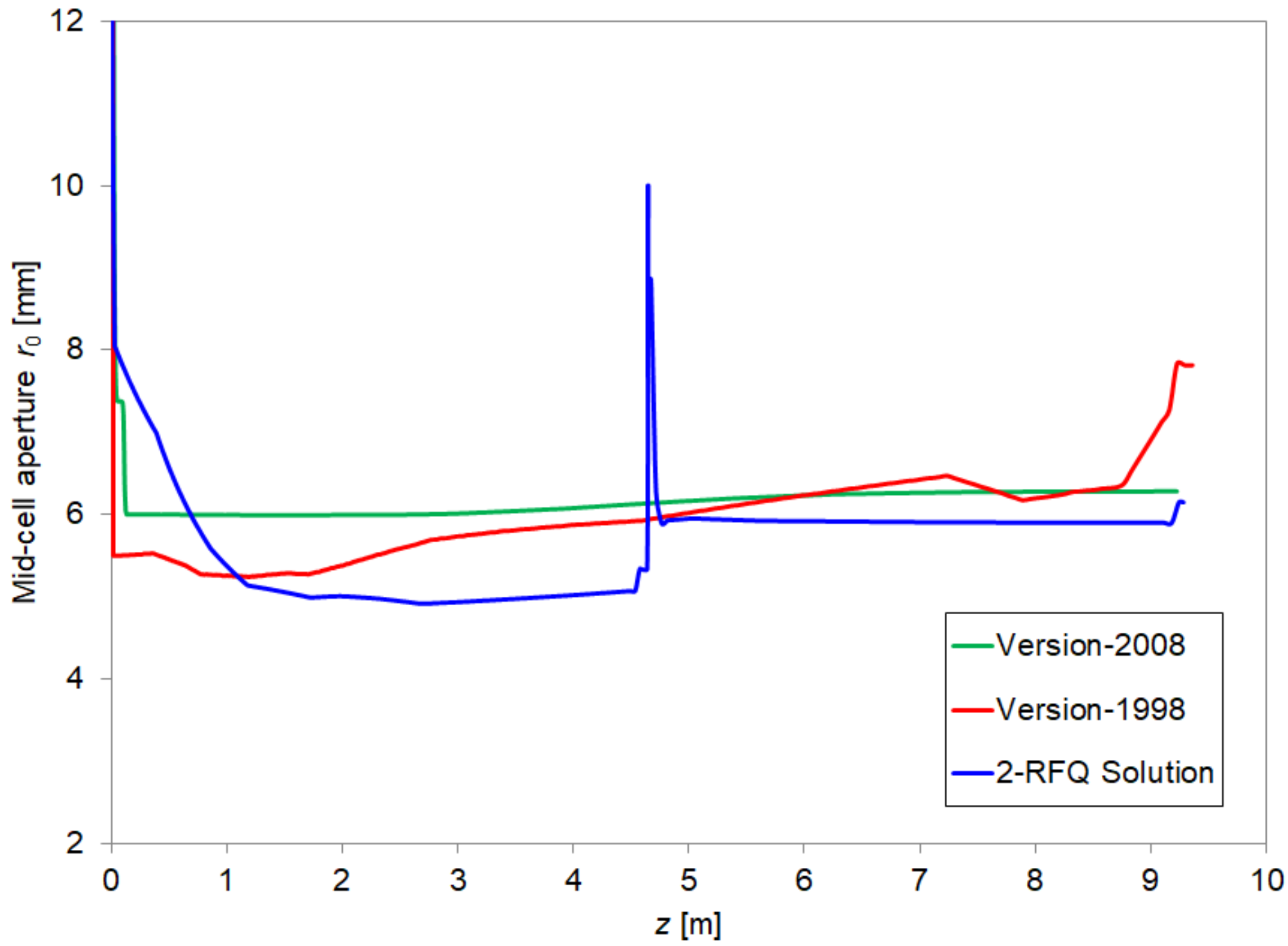

FIG. 3. Evolution of the mid-cell aperture along the accelerating channel (compared with the Version-1998 and Version-2008 RFQs).

The beam dynamics simulation of the RFQs has been performed using the PARMTEQM code [17] developed by Los Alamos National Laboratory. The 20 emA $U^{4+}$ input beam has a 4D-Waterbag distribution, and its transverse emittance is 210 π mm mrad (unnormalized, total) or 0.076 π mm mrad (normalized, rms), the same value as that used by the Version-2008 RFQ. The transverse beam envelopes for all particles as well as for 95% of the beam are shown in Fig. 4, where the black curves represent the mid-cell aperture of the RFQs and the beam hole aperture radius of the RF shielding wall between the RFQs. The total transmission of the linac is 96.05% and most losses happened between $z$ = 1 m and $z$ = 2 m.

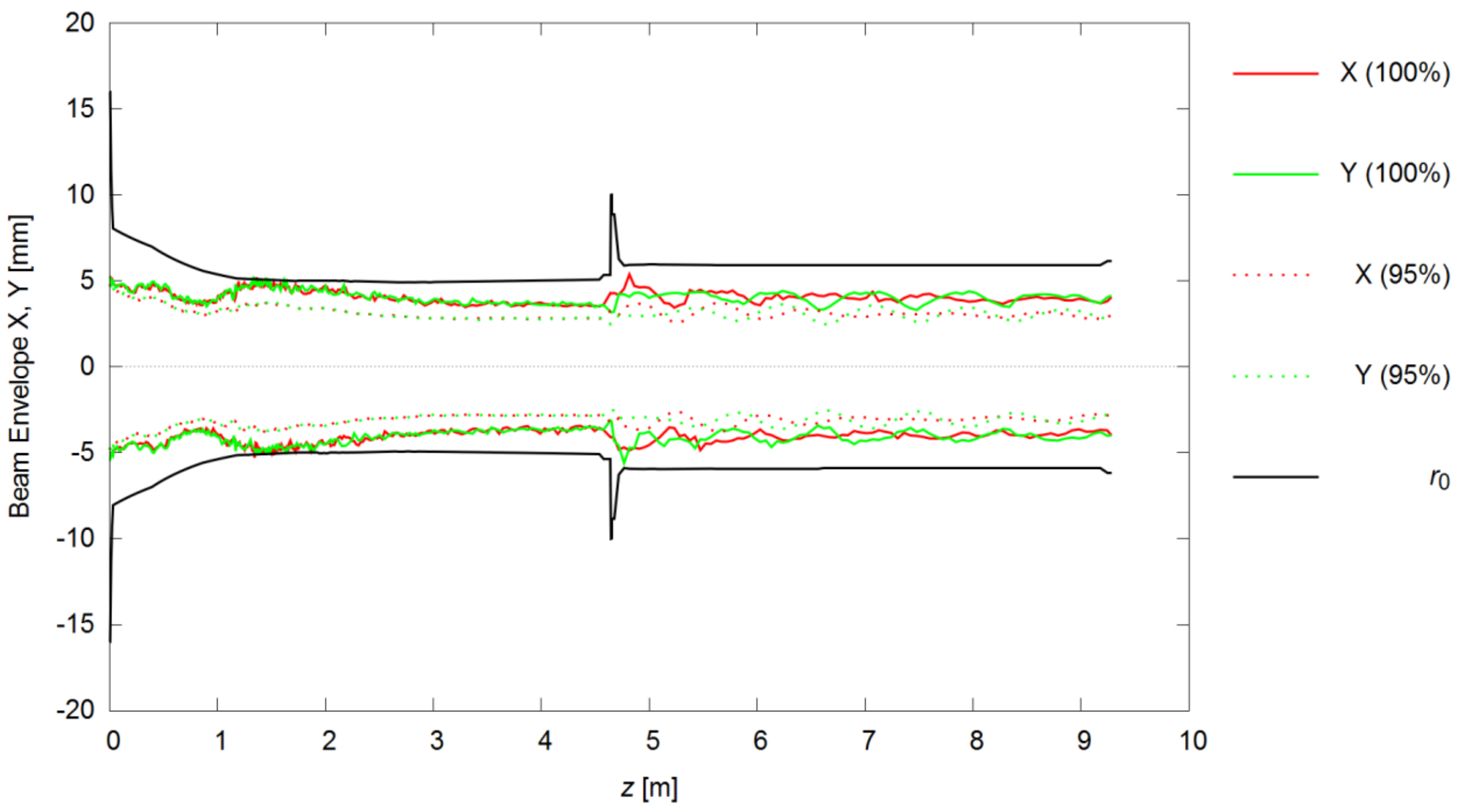

FIG. 4. Transverse beam envelopes for all particles as well as for 95% of the beam.

In Fig. 5, the transverse emittances are relatively constant as function of the longitudinal position, while the longitudinal emittance for all particles increases between $z$ = 2 m and $z$ = 7 m, but the longitudinal emittance curve for 99% of the beam indicates that this increase is due to only ≤1% of halo particles. It can be also seen that the design approach trying to hold $\frac{\varepsilon_l}{\varepsilon_t} = 1.0$ successfully maintained the beam quality along the whole linac.

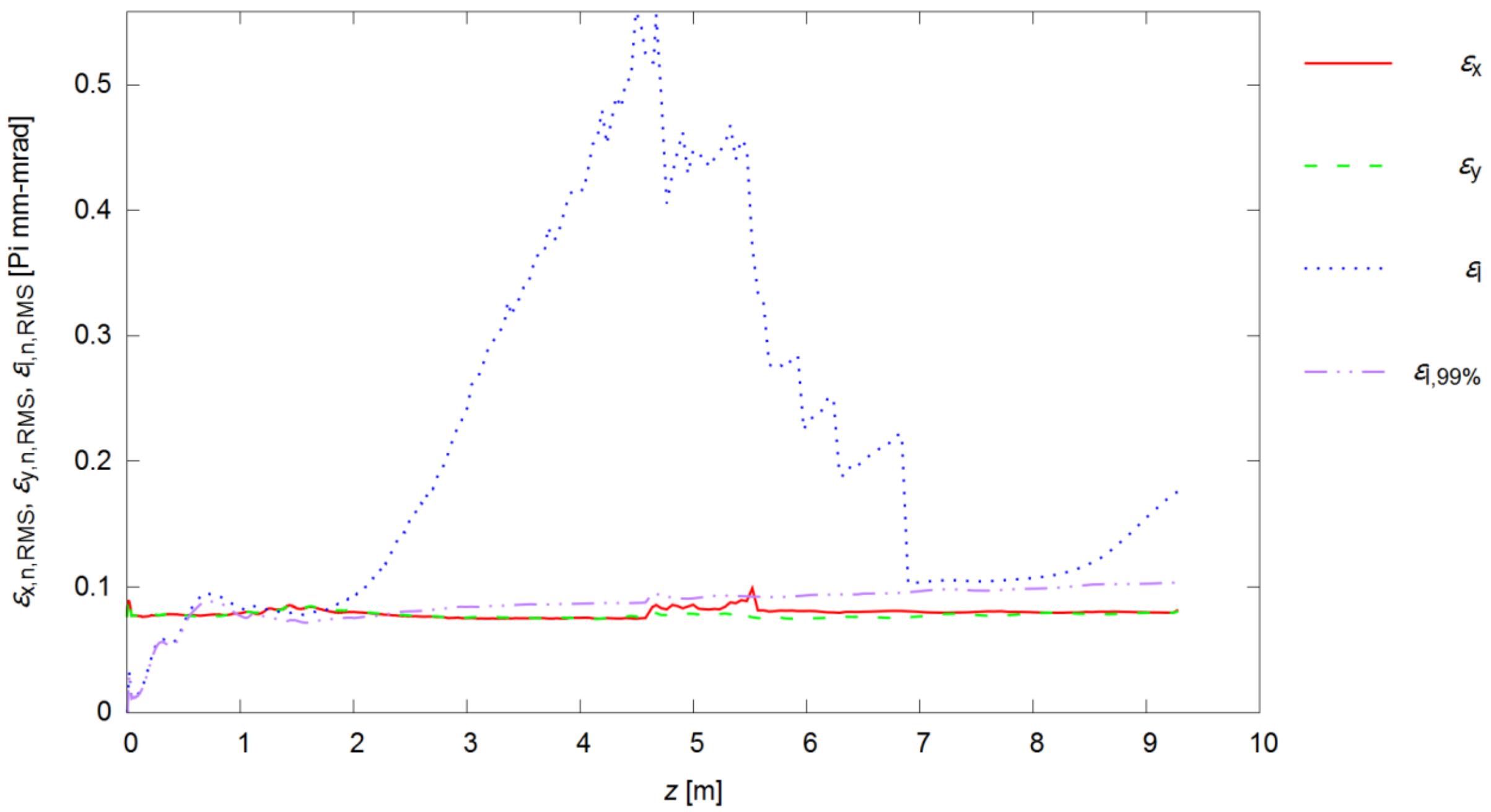

FIG. 5. Emittance evolutions along the linac (red: *x* plane, green: *y* plane, blue: longitudinal plane for all particles, purple: longitudinal plane for 99% of the beam).

The output distribution of the RFQ 1 (see the top graphs of Fig. 6) shows that the introduced transition cell has successfully provided $\alpha_{\mathrm{Twiss}} \cong 0$ transverse phase-space ellipses for entering into the RFQ 2. The beam distribution at the end of the whole linac (see the bottom graphs of Fig. 6) is still very concentrated. Its phase spread and energy spread for the main beam are well within ± 30° and ± 1%, respectively.

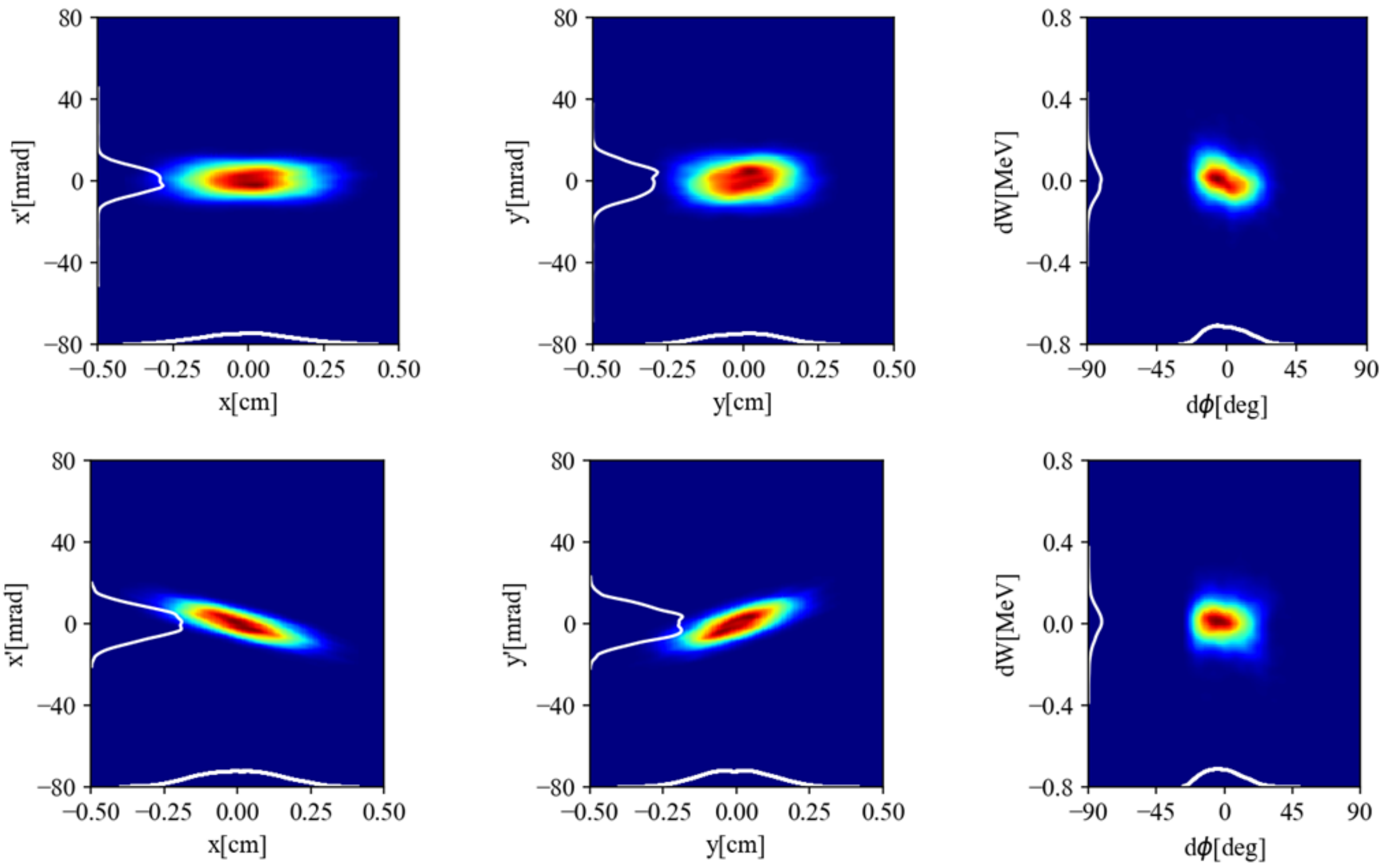

FIG. 6. Particle distributions at the RFQ 1 output (top graphs) and at the RFQ 2 output (bottom graphs).

The above results are based on the 2-RFQ solution with a short drift between the two RFQ cavities (the drift length $d$ = 1 cm, and the distance between two sets of electrodes is 3.1 cm). A batch of simulations have been performed by varying $d$ from 1 cm to 11 cm with an interval of 2 cm. Fig. 7 shows that the beam transmission is still ~ 90% when $d$ = 8 cm.

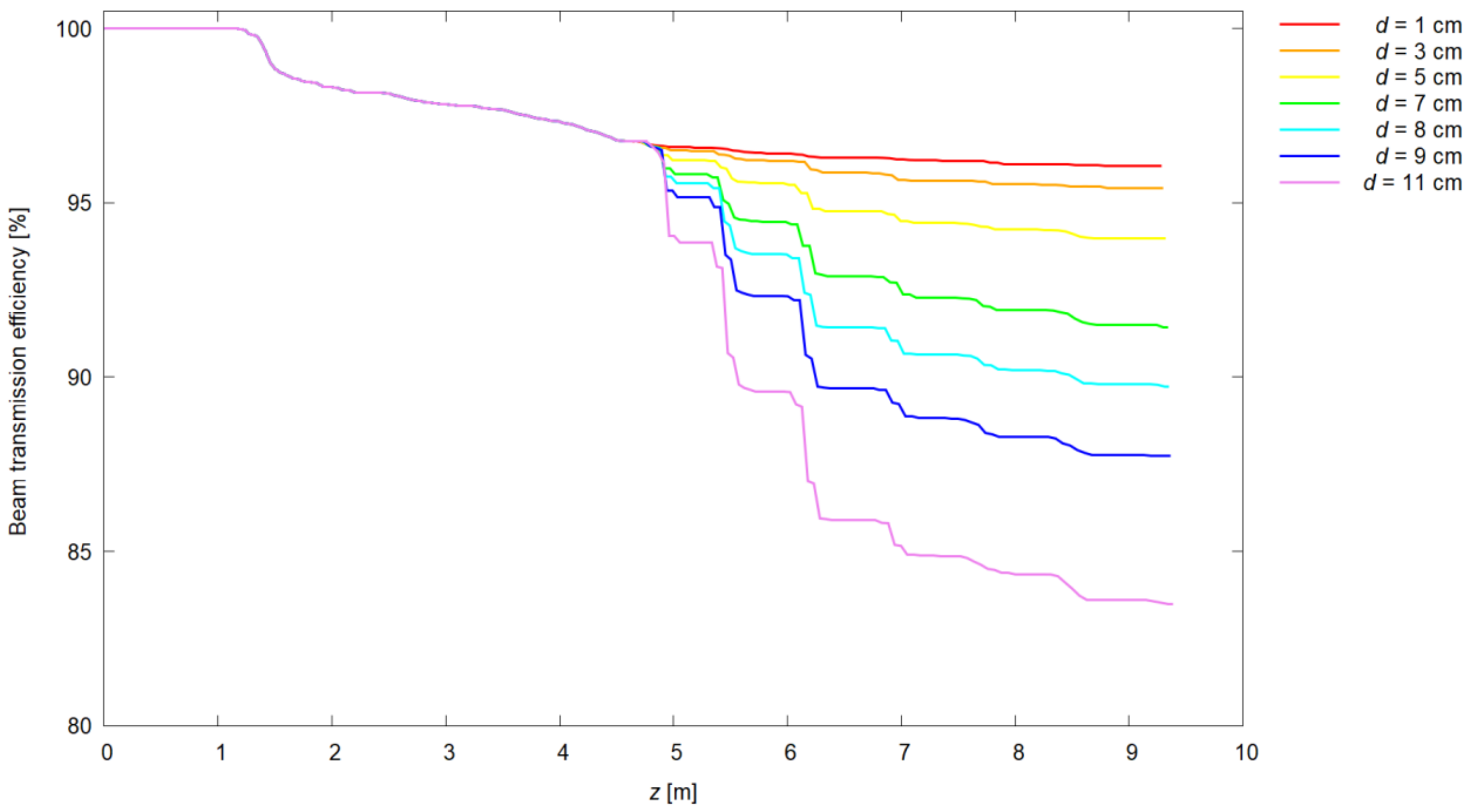

FIG. 7. Beam transmission curves for different drift lengths ($d$ = 1 – 11 cm).

Based on this result, different possible schemes can be applied to connect the two RFQ cavities whilst obtaining still reasonable transmission. One method is using a 1 – 5 cm thick wall to separate the RF fields of the two RFQs (see Fig. 8). Actually, it is also possible to have two individual end plates if $d \geq 4$ cm.

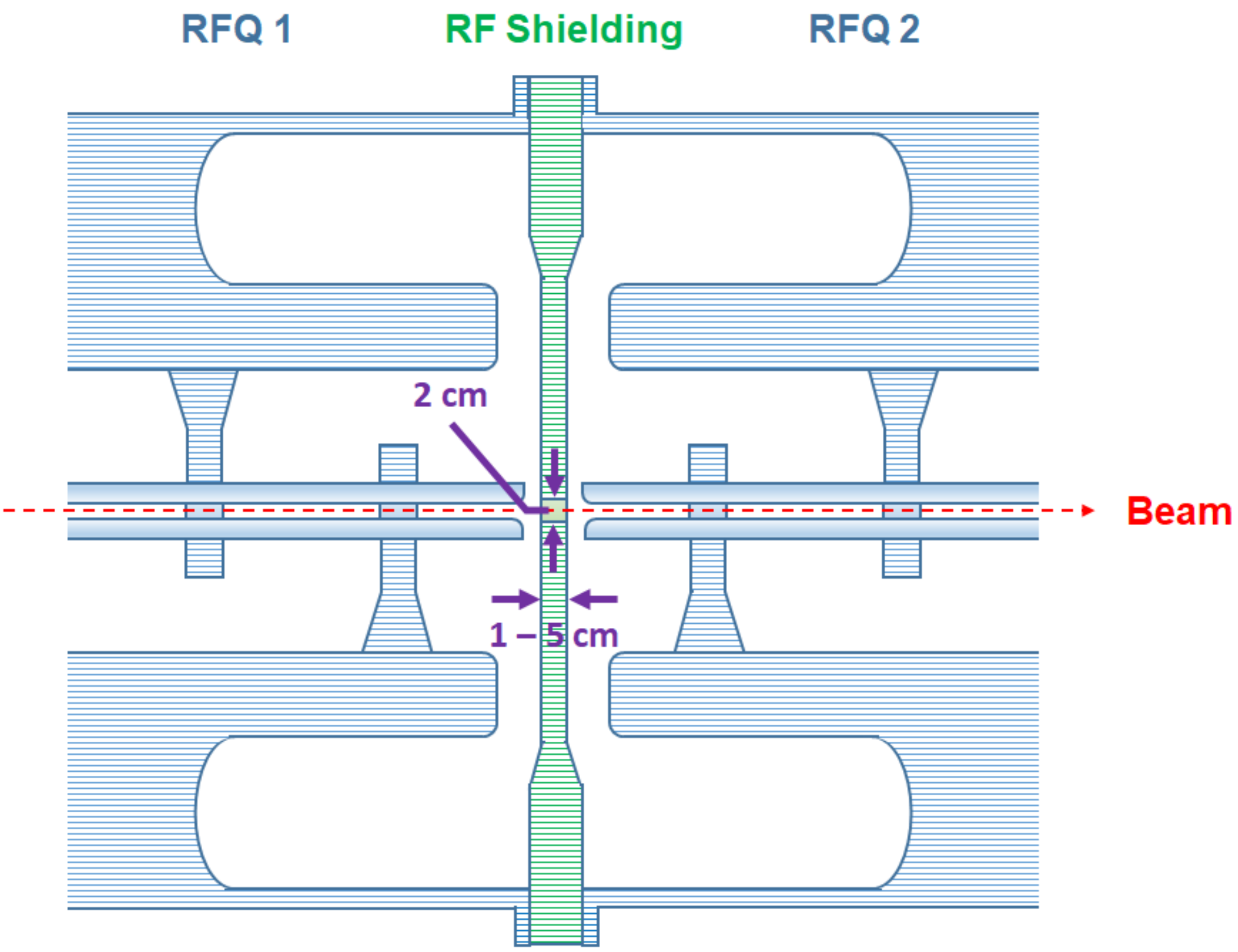

FIG. 8. Schematic plot for the connection of two RFQs by an RF shielding wall (the distance between two sets of electrodes excluding the RF shielding wall is 2.1 cm).

In case of 5 cm ≤ $d$ ≤ 8 cm, one beam diagnostic element e.g. an AC Current Transformer (ACCT) can be introduced into this drift space and integrated with the end plate as shown in Fig. 9. The beam hole aperture diameter at the two ends of the added element can be held as 2 cm for good RF shielding and then enlarged to 4 cm in the middle.

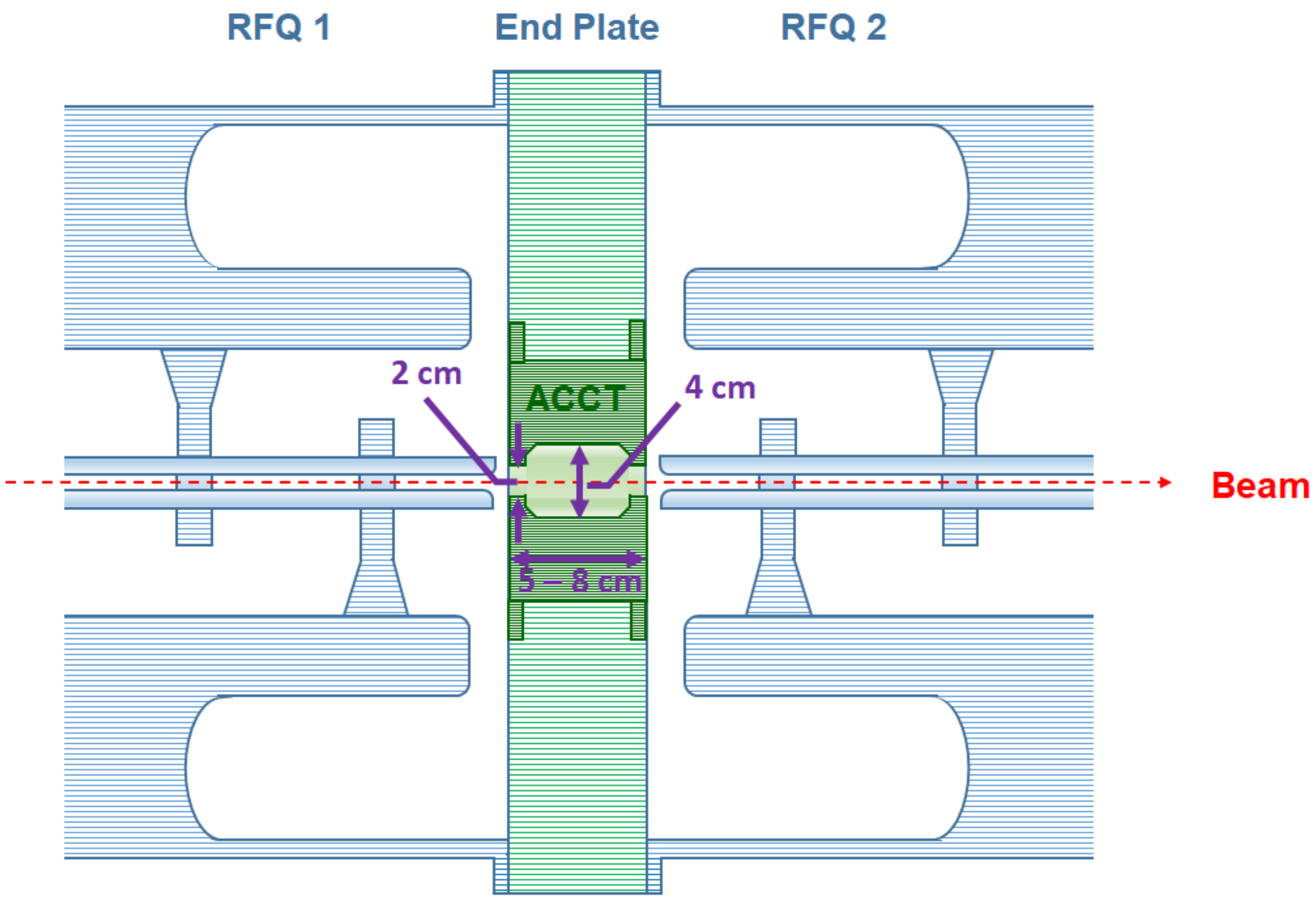

FIG. 9. Schematic plot of the connection between two RFQs for 5 cm ≤ $d$ ≤ 8 cm.

## III. 2-RFQ SOLUTION WITH MEBT BETWEEN CAVITIES

To allow more inter-tank elements e.g. XY-steerer and Beam Position Monitor (BPM) for a better beam matching and tuning in the operation, the 2-RFQ solution has been further extended by including an MEBT section. As shown in Fig. 10, the MEBT section consists of two 65 cm long triplets and one 3-gap rebuncher cavity. At 36.136 MHz, the Spiral structure [18] is a good candidate for the rebuncher. The main design parameters of the MEBT section are listed in Table II.

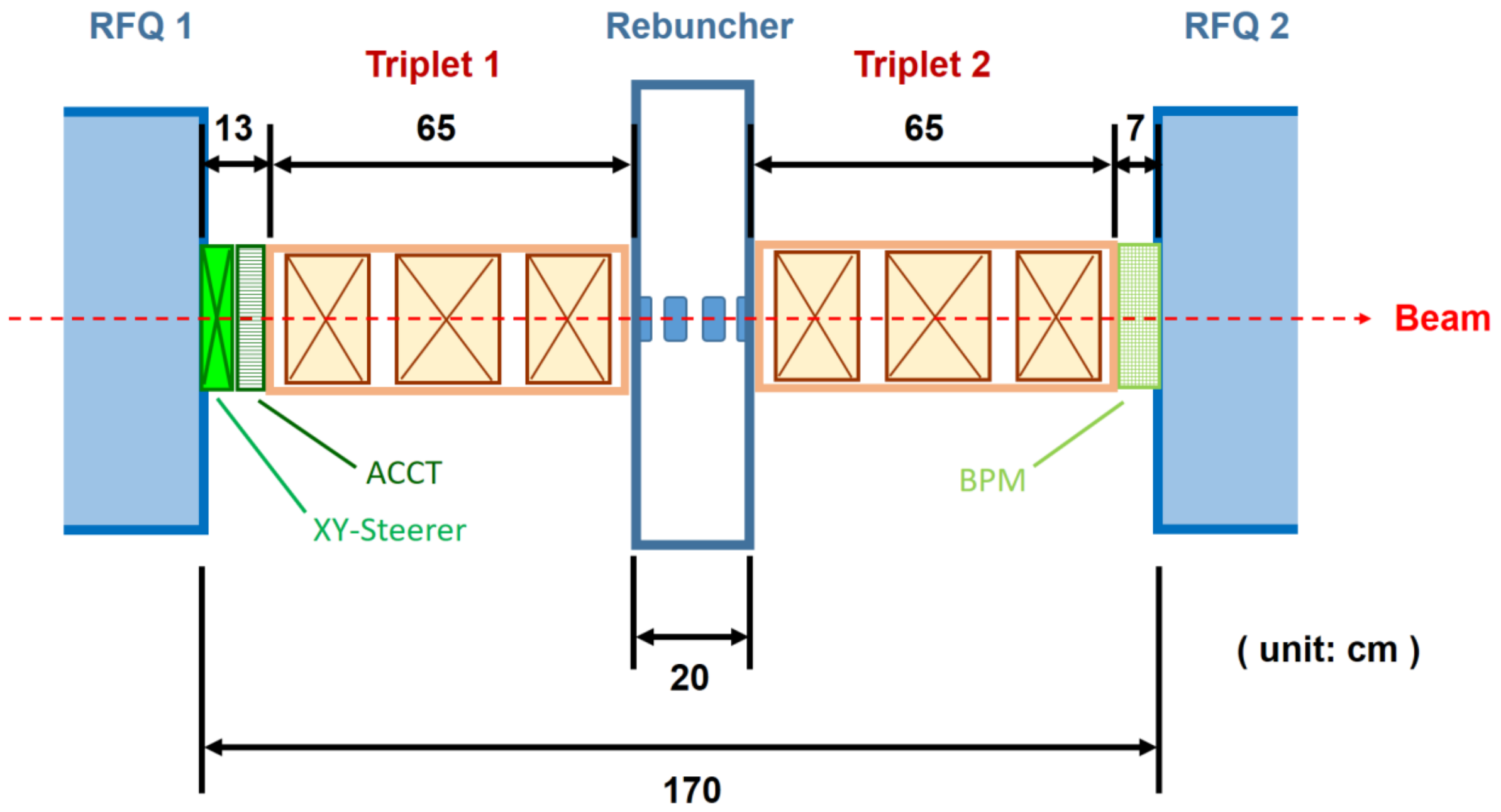

FIG. 10. Schematic plot of the connection between two RFQs with an MEBT.

TABLE II. Main design parameters of the MEBT section.

| | Triplet 1 | Rebuncher | Triplet 2 |
|---|---|---|---|
| Quadrupole length [cm] | 16 / 20 / 16 | | 16 / 20 / 16 |
| Quadrupole inner diameter [cm] | 3 | | 3 |
| Quadrupole pole-tip field [T] | $\leq 1.14$ | | $\leq 1.17$ |
| Resonant frequency [MHz] | | 36.136 | |
| Period length [cm] | | 4.45 | |
| Effective gap voltage [kV] | | 68 | |
| Tube inner diameter [cm] | | 2.4 | |

In this solution, the RFQ 1 is same as that mentioned in Section II due to its good performance. The MEBT has been designed in a way that its output distribution should be close to the RFQ 1 output distribution so that the matching to the following RFQ cavity will be as smooth as possible. The RFQ 2 uses also the same design, except the starting part has been slightly adapted according to the differences brought by the MEBT.

Using the RFQ 1 output distribution (see the top graphs of Fig. 6) as the input distribution for the MEBT section, the DYNAC code [19] originally developed by CERN has been adopted for the beam dynamics simulation and the LORASR code [20] has been taken for a benchmarking. Both codes gave very similar results. Although the magnetic focusing for such a very low beta and very heavy ion beam is difficult, a 94.5% beam transmission and good beam quality have been achieved. The transverse beam envelopes (see Fig. 11) have only some obvious oscillations in the MEBT part, but in the RFQ cavities they are still very similar to those of the "2-RFQ with drift" solution.

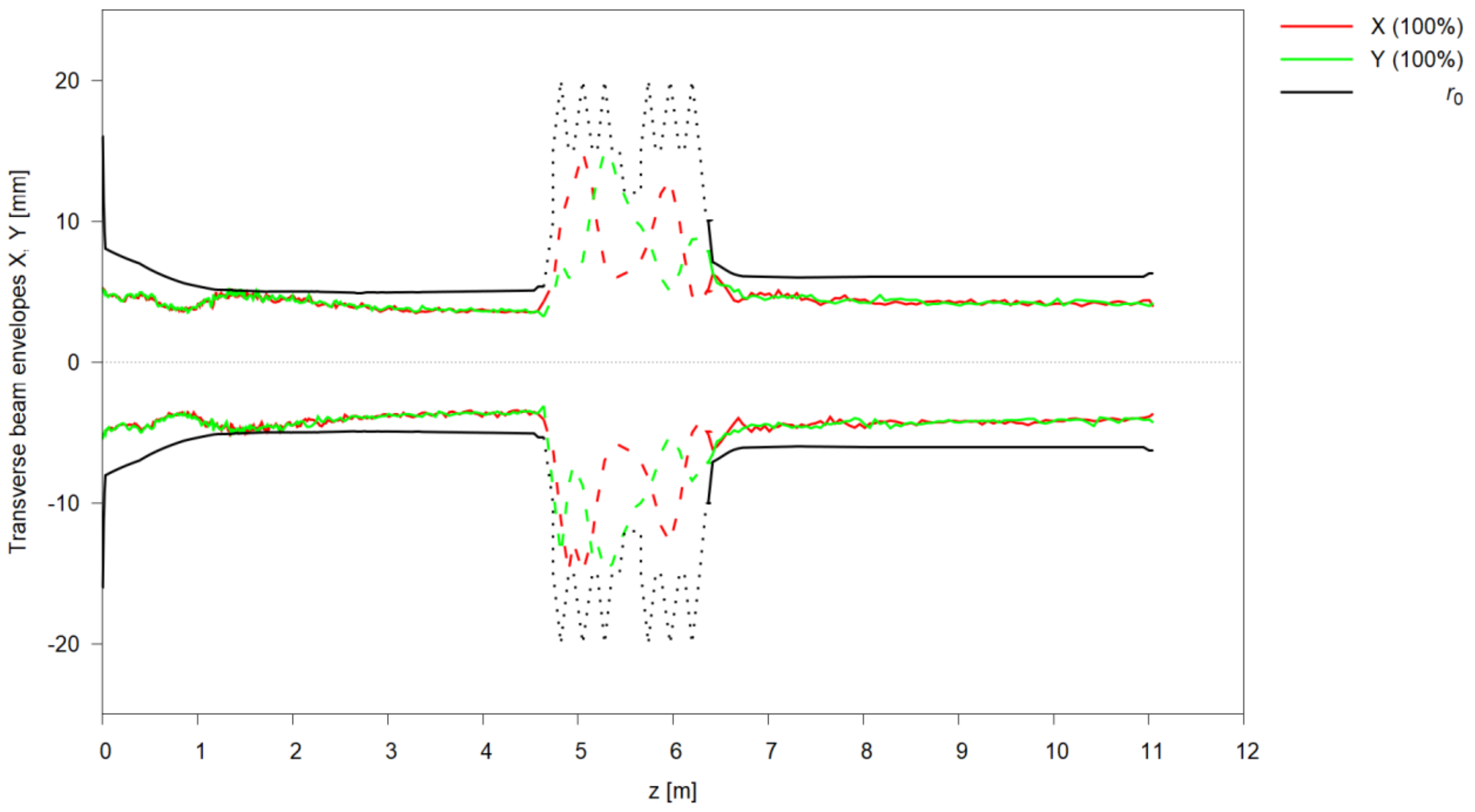

FIG. 11. Transverse beam envelopes for all particles along the linac (the solid lines are for the RFQs and the dashed lines are for the MEBT section, respectively).

Fig. 12 shows the transverse emittances are still relatively constant along the accelerating channel, despite the presence of the 1.7 m long MEBT. The longitudinal emittance has a little more growth with the new solution, but the 99%-emittance curve is still flat.

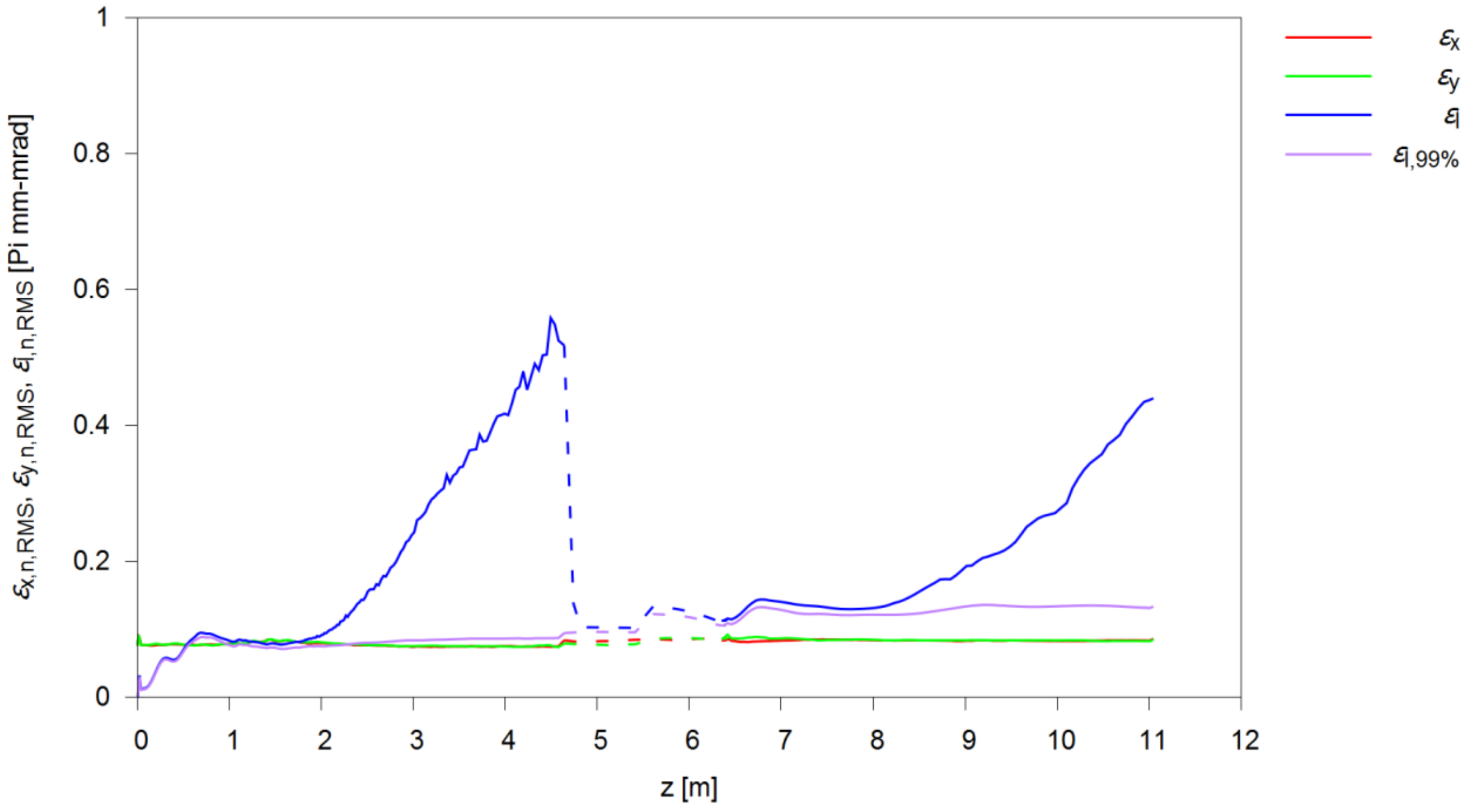

FIG. 12. Emittance evolutions along the linac with an MEBT (the solid lines for the RFQs, the dash lines for the MEBT section, the purple curve for 99% of the beam, respectively).

Fig. 13 shows the particle distributions at the MEBT exit and the whole linac exit. They are still comparable to the distributions shown in FIG. 6.

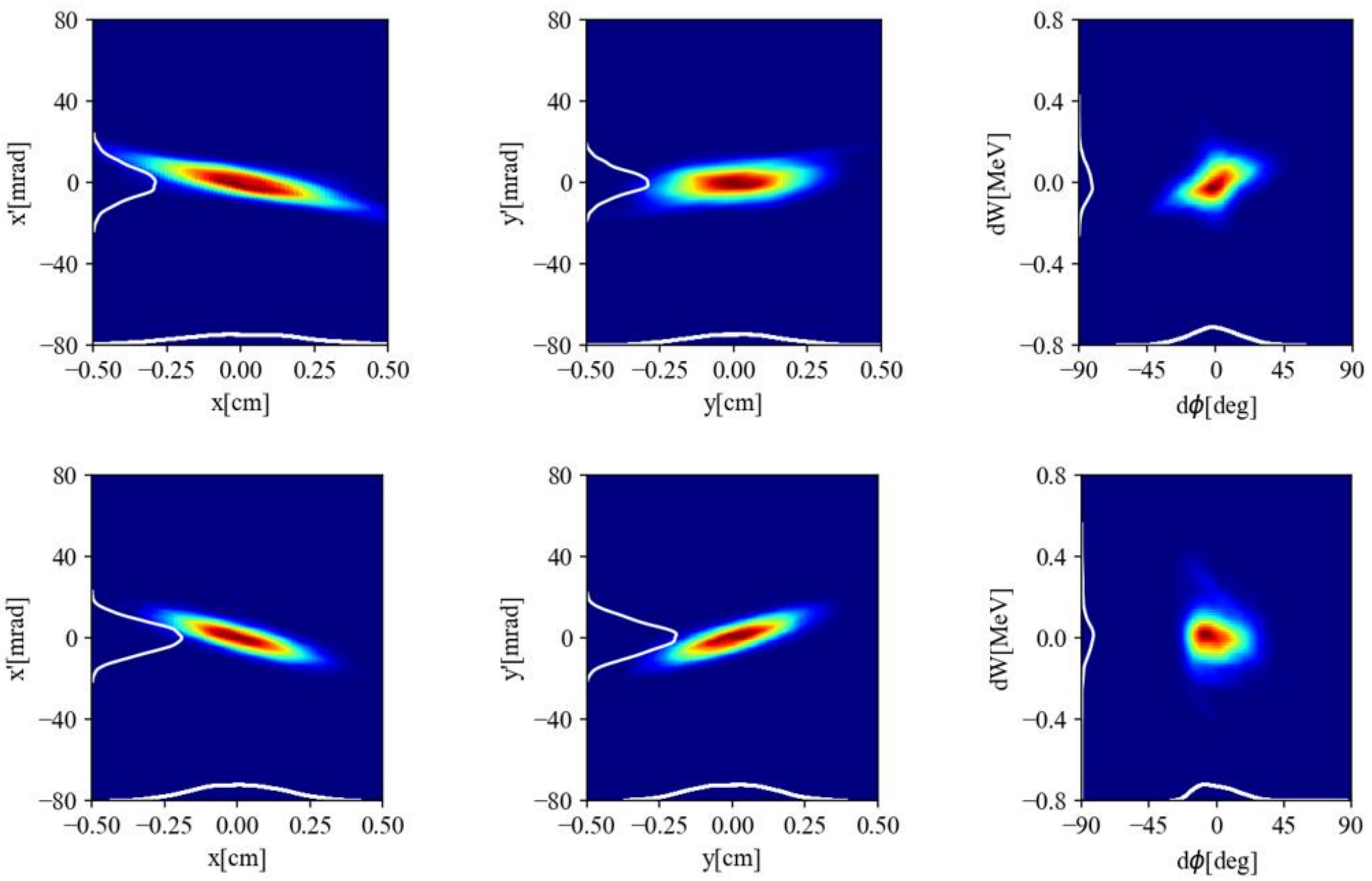

FIG. 13. Particle distributions at the MEBT output (top graphs) and at the RFQ 2 output (bottom graphs).

## IV. 3-RFQ SOLUTION WITH DRIFT SPACE BETWEEN CAVITIES

Finally, a design using 3 RFQs with drifts in between has been made. The transition energy values between the cavities have been chosen as 32.6 AkeV and 78.5 AkeV, respectively. Due to the different output energy, the RFQ 1 design has been modified, but still following the $\frac{\varepsilon_l}{\varepsilon_t} = 1.0$ design approach and the concept to have $\alpha_{\mathrm{Twiss}} \cong 0$ output transverse phase-space ellipses. The main design parameters of the new linac are shown in Fig. 14, where the RFQ 1, the RFQ 2, and the RFQ 3 are 3.5 m, 2.9 m and 2.9 m long, respectively. For this plot, the drift length between the cavities is 1 cm. Generally speaking, the parameters are varying in a similar way to those in the 2-RFQ case (see Fig. 2).

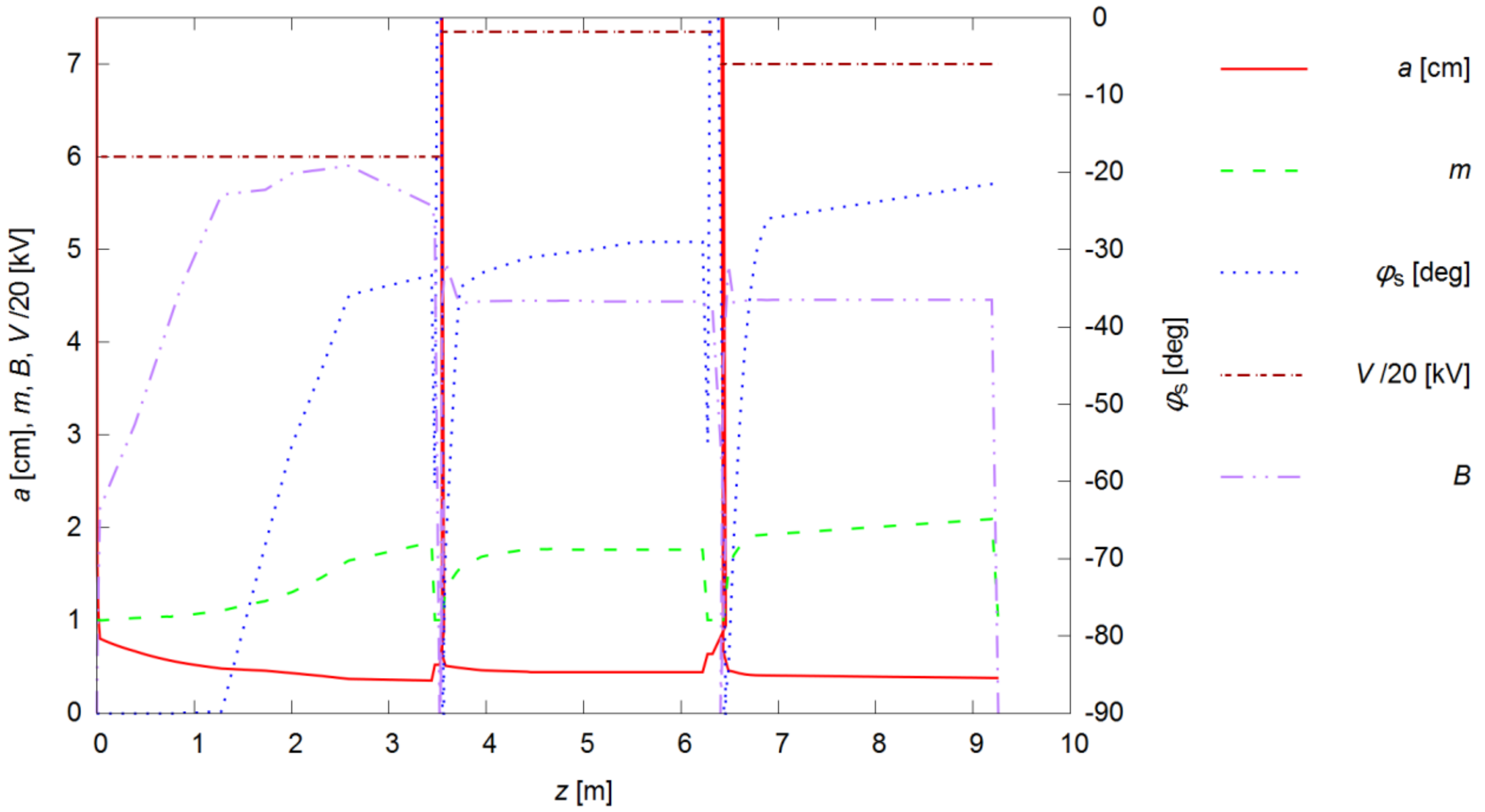

FIG. 14. Main design parameters of the “3-RFQ with drift” solution.

As the RFQ 1 stops at even lower output energy, the beam dynamics of this linac is more challenging due to the space charge. Nevertheless, the total length is only 3 cm longer and the total transmission is only 1.6% lower, compared with the “2-RFQ with drift” case. Further simulations show that the transmission is still ~ 90% when *d* is prolonged to 6 cm (without modification and optimization of the RFQ 2 and the RFQ 3). A comparison of the main parameters between this solution and the 2-RFQ solutions is given in Table III.

TABLE III. A comparison of the 2-RFQ and 3-RFQ solutions.

| | **2-RFQ w. drift** | | **2-RFQ w. MEBT** | **3-RFQ w. drift** | |
|---|---|---|---|---|---|
| Number of RFQs | 2 | 2 | 2 | 3 | 3 |
| Transition between cavities | drift | drift | MEBT | drift | drift |
| Transition length *d* [cm] | 1 | 8 | 170 | 1 (each) | 6 (each) |
| Maximum surface E-field $E_{s,max}$ [MV/m] | 30.9 | 30.9 | 30.9 | 30.9 | 30.9 |
| Inter-vane voltage *V* [kV] | 120 / 140 | | 120 / 140 | 120 / 147 / 140 | |
| Output emit. [π mm mrad] | | | | | |
| $\varepsilon_{x, n., rms, 100\%}$ | 0.082 | 0.092 | 0.086 | 0.086 | 0.094 |
| $\varepsilon_{x, n., rms, 99\%}$ | 0.079 | 0.088 | 0.083 | 0.083 | 0.090 |
| Output emit. [π mm mrad] | | | | | |
| $\varepsilon_{y, n., rms, 100\%}$ | 0.081 | 0.087 | 0.085 | 0.084 | 0.094 |
| $\varepsilon_{y, n., rms, 99\%}$ | 0.078 | 0.084 | 0.082 | 0.081 | 0.090 |
| Output emit. [π mm mrad] | | | | | |
| $\varepsilon_{l, n., rms, 100\%}$ | 0.176 | 0.350 | 0.440 | 0.464 | 1.142 |
| $\varepsilon_{l, n., rms, 99\%}$ | 0.106 | 0.131 | 0.134 | 0.117 | 0.144 |
| Total length *L* [m] | 9.28 | 9.35 | 11.05 | 9.31 | 9.41 |
| Total beam transmission *T* [%] | 96.1 | 89.7 | 94.5 | 94.5 | 89.3 |

From Table III, one can see that: 1) the maximum surface electric field for all these solutions is 30.9 MV/m, lower than that of both the Version-1998 RFQ and the Version-2008 RFQ; 2) the inter-vane voltage for the solutions is between 120 and 147 kV, lower than 155 kV used by the Version-2008 RFQ. These indicate that the new solutions can support a more reliable and more efficient (the power is proportional to $V^2$) operation.

In addition, the end-to-end evolution curves of synchronous energy and beam transmission efficiency are plotted in Fig. 15 and Fig. 16, respectively, for the three reference cases. Generally speaking, they are very comparable, if the MEBT is skipped.

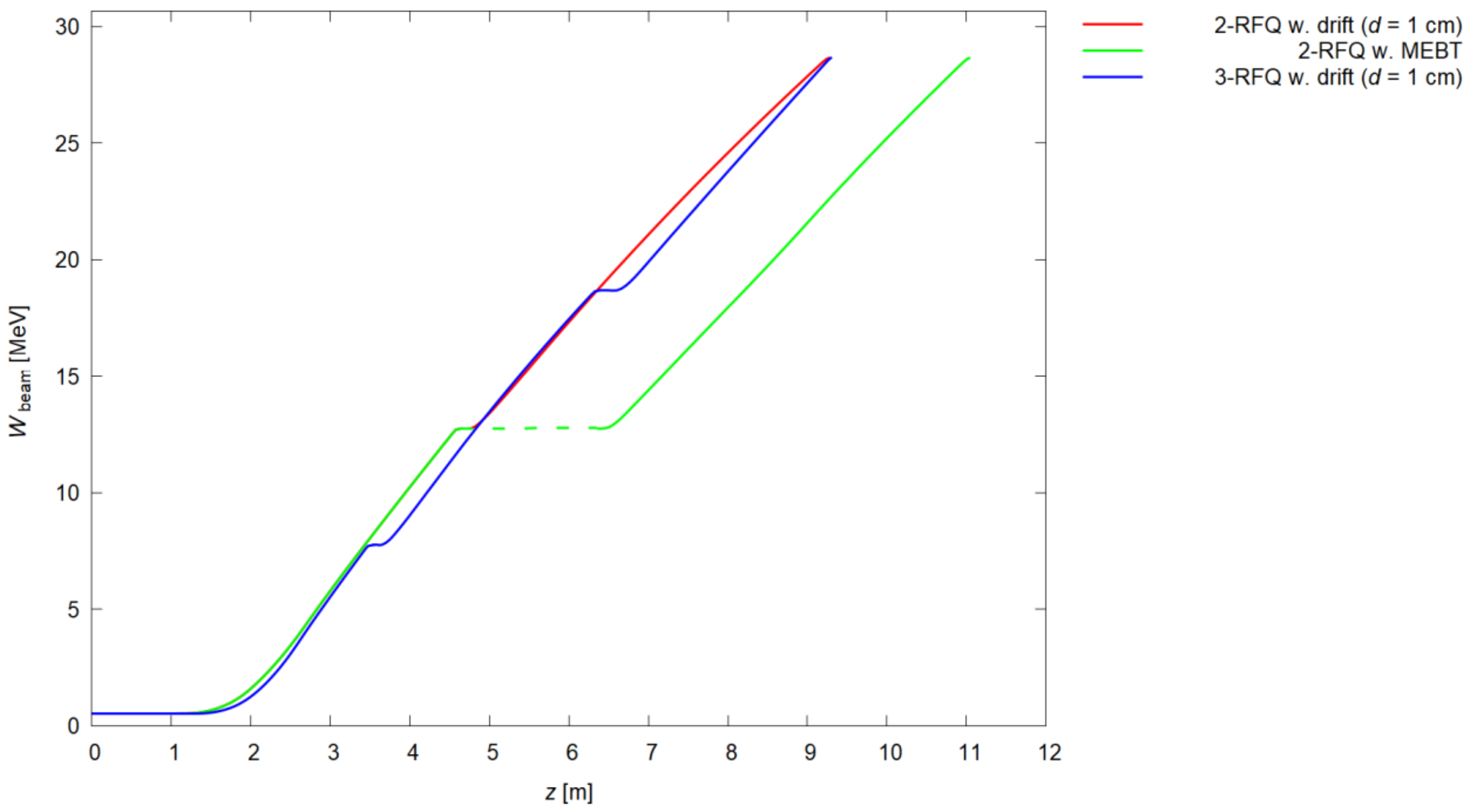

Fig. 15. Evolution of the synchronous energy for the three reference cases.

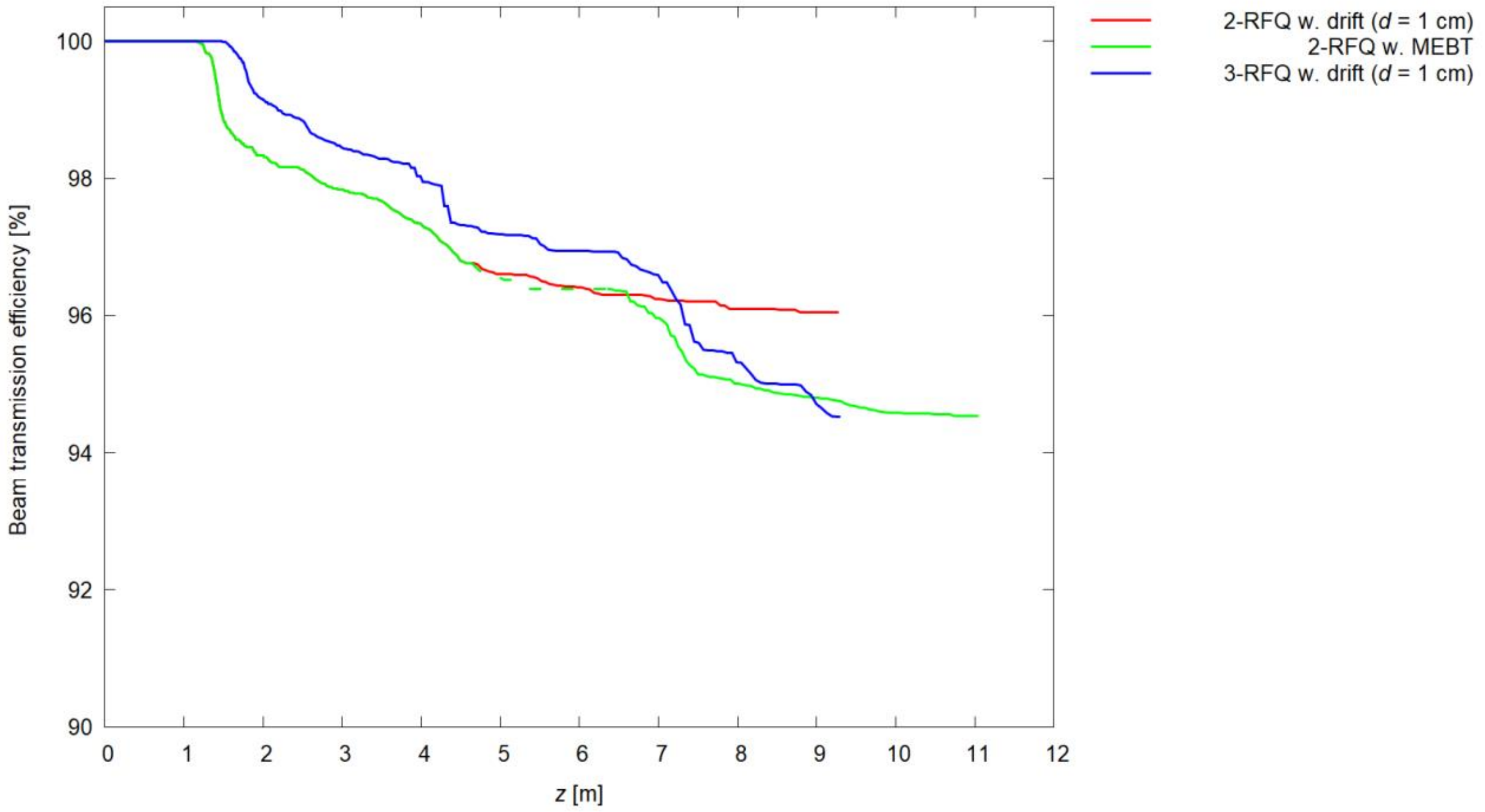

Fig. 16. Evolution of the beam transmission for the three reference cases.

## V. CONCLUSIONS

Different scenarios using multiple shorter and independent cavities to realize a longer than 9 m, low energy, and high current RFQ accelerator have been investigated. Compared to the single RFQ design, a solution using shorter cavities has the following advantages:

- It improves the longitudinal field stability and eases the RF tuning.
- One can use smaller and low-cost power amplifiers.
- It allows designing the different parts of the long accelerating channel individually and more efficiently. One can adapt the cavities to the changing beam situation along the beam line. For example, constant but different $r_0$ values can be applied for different cavities and also the inter-vane voltage of each cavity can be chosen more flexibly and reasonably.
- It supports to add beam diagnostics and knobs between cavities for a better beam matching and tuning in the operation.

Different from the resonant coupling approach that divides a long RFQ into several segments with short (a few mm long) gaps as transitions, which has only one radial matching section and one fringe field section, the multi-cavity approach used by this study has multiple radial matching sections and exit fringe field sections. Each cavity has to be treated separately, not only from an RF point of view but also in the sense of beam dynamics.

To solve the beam matching problem between cavities as well as to reach good beam quality and high transmission also at high current, the design of the first RFQ cavity is always most difficult. One key task for the RFQ 1 is to bunch the input beam at low energy, so it is very important to adopt the $\frac{\varepsilon_l}{\varepsilon_t} = 1.0$ design approach which can minimize the emittance transfer between the longitudinal and transverse planes and reduce beam instability. In addition, providing $\alpha_{\text{Twiss}} \cong 0$ transverse phase-space ellipses to the downstream cavities can allow a smooth matching and help maintaining the beam quality as well.

Benefiting from these special methods, the new multi-cavity approach developed by this study can not only improve the RF performance of the HSI RFQ e.g. larger mode separation, easier flatness tuning, lower sparking risk, and less power consumption, but also reach high beam transmission and good beam quality with a comparable total length.

## ACKNOWLEDGEMENTS

Special thanks go to R. Tiede, T. Sieber, H. Vormann, and A. Bechtold for the valuable discussions with respect to the two constructed HSI RFQs and the realistic dimensions for possible inter-tank elements.